\documentclass{article}

\usepackage{arxiv}

\usepackage[utf8]{inputenc} 
\usepackage[T1]{fontenc}    
\usepackage{hyperref}       
\usepackage{url}            
\usepackage{booktabs}       
\usepackage{amsfonts}       
\usepackage{nicefrac}       
\usepackage{microtype}      
\usepackage{lipsum}		
\usepackage{graphicx}
\usepackage[numbers]{natbib}
\usepackage{doi}

\usepackage{multirow}%
\usepackage{comment}

\usepackage{amsmath,amssymb,amsfonts}%
\usepackage{amsthm}%
\usepackage{mathrsfs}%
\usepackage{newtxtext,newtxmath}  
\usepackage{lineno,hyperref}
\usepackage[title]{appendix}%
\usepackage{xcolor}%
\usepackage{textcomp}%
\usepackage{manyfoot}%
\usepackage{longtable}
\usepackage{bm}
\usepackage{array}
\usepackage{subcaption}
\usepackage{color}
\usepackage{algorithm}%
\usepackage{algorithmicx}%
\usepackage{algpseudocode}%
\usepackage{listings}%

\newcommand \modificacion[1]        {{\color{black}#1}} 
\newcommand \modificaciondos[1]        {{\color{black}#1}} %

\title{A Memory-Efficient Distributed Algorithm for Approximate Nearest Neighbour Search with Arbitrary Distances}

\author{ \href{https://orcid.org/0000-0003-2360-0394}{\includegraphics[scale=0.06]{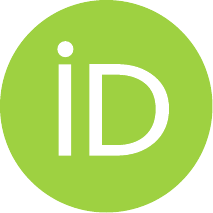}\hspace{1mm}Elena Garcia-Morato}\thanks{Corresponding author.} \\
	CETINIA-DSLAB\\
    Universidad Rey Juan Carlos\\
    Madrid, Spain\\
	\texttt{elena.garciamorato@urjc.es} \\
	\And
	\href{https://orcid.org/0000-0002-7539-8522}{\includegraphics[scale=0.06]{figures/orcid.pdf}\hspace{1mm}Maria Jesús Algar} \\
	CETINIA-DSLAB\\
    Universidad Rey Juan Carlos\\
    Madrid, Spain\\
	\texttt{mariajesus.algar@urjc.es} \\
    \And
	\href{https://orcid.org/0000-0002-9146-4067}{\includegraphics[scale=0.06]{figures/orcid.pdf}\hspace{1mm}Cesar Alfaro} \\
	CETINIA-DSLAB\\
    Universidad Rey Juan Carlos\\
    Madrid, Spain\\
	\texttt{cesar.alfaro@urjc.es} \\
    \And
	\href{https://orcid.org/0000-0003-0231-2051}{\includegraphics[scale=0.06]{figures/orcid.pdf}\hspace{1mm}Felipe Ortega} \\
	CETINIA-DSLAB\\
    Universidad Rey Juan Carlos\\
    Madrid, Spain\\
	\texttt{felipe.ortega@urjc.es} \\
    \And
	\href{https://orcid.org/0000-0001-6434-7263}{\includegraphics[scale=0.06]{figures/orcid.pdf}\hspace{1mm}Javier Gomez} \\
	CETINIA-DSLAB\\
    Universidad Rey Juan Carlos\\
    Madrid, Spain\\
	\texttt{javier.gomez@urjc.es} \\
    \And
	\href{https://orcid.org/0000-0003-1415-1961}{\includegraphics[scale=0.06]{figures/orcid.pdf}\hspace{1mm}Javier M. Moguerza} \\
	CETINIA-DSLAB\\
    Universidad Rey Juan Carlos\\
    Madrid, Spain\\
	\texttt{javier.moguerza@urjc.es} \\
}

\date{}

\renewcommand{\headeright}{}
\renewcommand{\undertitle}{Published in IEEE Open Journal of the Computer Society on 2 October 2026. \\The version of record is available at
\href{https://doi.org/10.1109/OJCS.2026.3739651}
{doi: 10.1109/OJCS.2026.3739651}.}

\renewcommand{\shorttitle}{A Memory-Efficient Distributed Algorithm for Approximate Nearest Neighbour Search with Arbitrary Distances}

\hypersetup{
pdftitle={A Memory-Efficient Distributed Algorithm for Approximate Nearest Neighbour Search with Arbitrary Distances},
pdfsubject={q-bio.NC, q-bio.QM},
pdfauthor={David S.~Hippocampus, Elias D.~Striatum},
pdfkeywords={Approximate similarity search, arbitrary distance function},
}

\begin{document}
\maketitle

\begin{abstract}
Approximate nearest neighbour (ANN) search has become a central task in modern data-intensive applications, particularly when operating on large-scale, heterogeneous, or high-dimensional datasets. However, many existing ANN methods struggle in such scenarios, either because they rely on metric assumptions or because their indexing strategies are not well suited to distributed environments or to settings with constrained memory resources. This work introduces PDASC (Parametrizable Distributed Approximate Similarity Search with Clustering), a distributed ANN search algorithm \modificacion{whose index design jointly addresses three concerns commonly treated separately in the ANN literature: support for arbitrary dissimilarity functions, native distributed execution, and efficient operation under constrained memory environments, avoiding reliance on specialised hardware accelerators or costly memory resources.} PDASC builds a distributed hierarchical index that accommodates non-metric and domain-specific similarities while naturally partitioning indexing and search across multiple computing nodes, with a compact per-node memory footprint. By preserving locally informative neighbourhood structure, the proposed index \modificacion{reduces sensitivity to} practical manifestations of the curse of dimensionality in high-dimensional spaces. We analyse how the index structural parameters govern the trade-offs among recall, computational cost, and memory usage. Experimental evaluation across multiple benchmark datasets and distance functions shows that PDASC achieves competitive accuracy–efficiency trade-offs while consistently requiring lower per-node memory compared to state-of-the-art ANN methods.
\end{abstract}



\maketitle

\section{INTRODUCTION}
\label{intro}



Searching for the closest data objects to a given query according to a similarity 
measure, typically formalised as a distance function, is a core task in today’s data-driven landscape, commonly referred to as similarity search \cite{Chavez2001, Zezula2006}. This paradigm underpins many fields, including information retrieval, content-based multimedia search, machine learning, bioinformatics, and cybersecurity.

A particularly important instance of this paradigm is the $k$-nearest neighbours ($k$-NN) search problem, where we seek to retrieve the $k$ most similar objects to a given query according to the same dissimilarity function. A naive approach to $k$-NN search consists of computing the distance between the query and every element in the dataset, then selecting the $k$ elements with the smallest distances. While this exhaustive search guarantees exact results, it quickly becomes computationally prohibitive as the dataset grows in size or dimensionality. 
To address these limitations, Approximate Nearest Neighbour (ANN) search has emerged as a practical alternative, relaxing strict accuracy requirements in exchange for substantial efficiency gains.

Beyond classical high-dimensional challenges, modern data science introduces additional complexities, as datasets with heterogeneous features become increasingly prevalent \modificacion{and are often inherently distributed or too large to be processed on a single machine. Likewise, the growing cost of specialised hardware and high-memory infrastructures further complicates similarity search in large-scale environments. While individual methods address one or two of these aspects, few attempt to tackle all three within a unified framework. For instance, despite their strong performance in centralised settings, widely adopted approaches such as HNSW or IVF are not designed for distributed, memory-constrained deployments and provide limited or no native support for non-metric or domain-specific distance functions, leaving a practical gap in the ANN landscape.}



Building on these considerations, we propose PDASC (Parametrizable Distributed Approximate Similarity Search with Clustering), a fully distributable ANN search algorithm capable of operating under a wide range of distance functions \modificacion{and memory-constrained architectures, without hardware acceleration strategies such as GPUs or TPUs or costly memory resources. A  distinctive characteristic is its index construction strategy, which relies on a tailored application of clustering techniques that preserves locally informative neighbourhood structure, thereby reducing sensitivity to practical manifestations of the curse of dimensionality in high-dimensional spaces.}


In summary, the main contributions of this research are:

\begin{itemize}
    \vspace{-0.6em}
    \item \modificacion{Proposing PDASC, a hierarchical, clustering-based ANN search method integrating support for arbitrary distance functions, native distributed execution, and a compact per-node memory footprint within a unified design.}
    
    \item \modificacion{Providing a structural sensitivity analysis of the PDASC index, characterising how different index topologies and distributed configurations influence recall, computational cost, and memory usage under different search conditions. The analysis identifies effective operating ranges that yield competitive accuracy--efficiency trade-offs across diverse datasets and distance functions.}

    \item \modificacion{Conducting an extensive empirical evaluation of PDASC across multiple benchmark datasets and distance functions, including non-metric and domain-specific measures, and comparing its performance against state-of-the-art ANN algorithms. Results demonstrate its robustness in distributed environments and competitive accuracy--efficiency trade-offs compared to established methods, while maintaining consistently low per-node memory requirements.}
        
\end{itemize}


The paper is organised as follows. Section~\ref{sec:related-work} formalises our problem statement and related work. Section~\ref{sec:proposed_algorithm} describes the PDASC algorithm, and Section~\ref{sec:experimental_evaluation} details the empirical evaluation conducted on benchmark datasets and presents the corresponding findings. Finally, Section~\ref{sec:conclusion} summarises the main contributions of this paper and discusses further directions for future work.

\section{BACKGROUND AND RELATED WORK}
\label{sec:related-work}


\subsection{SIMILARITY SEARCH}
\label{subsec:similarity-search}

Similarity search can be formally defined over a dataset of objects $\mathbb{X}$, each represented as a feature vector in a multidimensional space, together with a dissimilarity function $\delta: \mathbb{X} \times \mathbb{X} \to \mathbb{R}$ that quantifies how different two objects $a,b \in \mathbb{X}$ are by assigning a numerical score $\delta(a,b)$. Larger values of $\delta(a,b)$ indicate greater dissimilarity, whereas smaller values suggest more similar objects. Given a query object $q \in \mathbb{X}$, the goal is to identify the object $o \in \mathbb{X}$ or, more generally, a subset $S \subseteq \mathbb{X}$, minimising $\delta(q,o)$.

In particular, the $k$-NN search retrieves the subset of $k$ objects in the database closest to the query $q$ according to the dissimilarity measure $\delta$, i.e., the $k$ elements $o_i \in \mathbb{X}$, where $i = 1, \ldots, k$, with the smallest values of $\delta(q,o_i)$.
The computational cost of similarity search queries primarily arises from comparing a query object with a large collection of high-dimensional feature vectors, implying numerous evaluations of the dissimilarity function  $\delta(a, b)$ \citep{Samet2006}. 

These queries can be addressed either in an \textit{exact} or \textit{approximate} manner. Exact methods guarantee the retrieval of the complete and correct result set, such as all true nearest neighbours 
, even at the cost of evaluating every distance. Approximate methods, in contrast, trade a small degree of accuracy for substantial gains in efficiency, either by reducing the number of distance computations (NDC) or by exploiting compact indexing structures.

\modificacion{Search tasks} in high-dimensional spaces are affected by phenomena such as the \textit{empty space problem}~\cite{Chavez2001} and \textit{measure concentration}~\cite{scott_thompson_1983}. As dimensionality increases, the volume of the space grows exponentially, whereas the point density decreases. As a consequence, distances become increasingly concentrated, with the relative difference between nearest and farthest neighbours shrinking and thereby reducing the discriminative power of distance-based proximity relationships, a phenomenon commonly referred to as the \textit{curse of dimensionality}~\citep{Bohm2001}. \modificacion{Although recent advances include approaches that improve the efficiency of exact $k$-NN search~\cite{Pan2020} and adaptive data structures that mitigate some effects of the curse of dimensionality in exact search~\cite{prior2025}, approximate methods continue to represent an attractive alternative in many large-scale high-dimensional settings, where small retrieval inaccuracies can often be exchanged for substantial gains in efficiency and scalability.}




\subsection{ANN SEARCH}
\label{subsec:ANN}

Following the exact $k$-NN formulation (see Section~\ref{subsec:similarity-search}), ANN search aims to retrieve the $k$ elements of $\mathbb{X}$ that best approximate the true nearest neighbours of $q$ under the dissimilarity function $\delta$, thereby relaxing strict accuracy requirements in exchange for significant efficiency gains.

ANN methods typically operate in two distinct phases: an index-building phase and a search phase. During the index-building phase, the dataset is mapped into a so-called index data structure tailored to enable efficient exploration of the search space. These index structures can be broadly classified into several families, including \modificacion{space-partitioning methods (encompassing tree-based and clustering-based approaches, among others)\cite{Wang23,vargas2019hcnng}}, proximity graph-based approaches \cite{MalkovY20,Wang2021}, hashing techniques \cite{mic2018binary,Zhao2023}, vector quantisation methods \cite{johnson2019billion,Xu2018}, permutation-driven approaches \cite{amato2014some}, and, more recently, learned and hybrid indices \cite{antol2021learned,Kraska2018learnedidx}. \modificacion{Orthogonally to the underlying index structure, recent research has also explored distributed ANN systems designed to scale similarity search across multiple computing nodes \cite{wang2021_milvus, deng2019pyramid, echihabi2021trends}.} 

\modificacion{From the search-strategy perspective}, ANN methods exploit the inherent structural properties of the index to constrain the portion of the search space to be examined, thus reducing the number of candidate elements to evaluate. Some approaches achieve this implicitly through their algorithmic design. For example, graph-based indices guide the search via greedy or heuristic traversals on a proximity graph. In contrast, hashing-based techniques exploit the distribution of collisions in the hash spaces. Others rely on explicit pruning rules to further limit the search space. In metric settings, several of these pruning rules are relaxed or heuristic variants of geometric strategies originally developed for exact search\modificaciondos{~\cite{Ciaccia1997}}. For instance, the triangle inequality has traditionally been employed to derive distance bounds from previously computed distances to reference elements, effectively excluding regions that cannot contain closer neighbours~\cite{Zezula2006}. However, such pruning strategies are not universally applicable: methods that operate with arbitrary or non-metric dissimilarity functions cannot reliably exploit geometric properties to guide the search. 

\subsection{NON-METRIC DISTANCE FUNCTIONS ON ANN SEARCH}
\label{subsec:dist-funcs}
Although approximate nearest-neighbour search in high-dimensional Euclidean spaces is also computationally expensive~\cite{indyk_approximate_1998, Beyer1999}, Euclidean distance remains widely used in indexing structures and search algorithms~\cite{Li2020ANN}, \modificacion{largely because many similarity search methods are built upon geometric properties derived from Euclidean spaces. However, this choice does not always align with the underlying semantics of complex datasets}. Indeed, many practical similarity measures used in domains such as text, vision, speech, and recommender systems (including cosine-based distances, Kullback-Leibler and Jensen-Shannon divergences, or learned similarity functions) are inherently non-metric and often capture domain-specific relationships more effectively than classical metric distances. Moreover, recent studies~\citep{ehsani_robust_2020, Aumuller2020} have shown that the choice of distance function can exert a significant impact on $k$-NN performance, \modificacion{since different dissimilarity measures may induce substantially different neighbourhood structures over the same dataset.}

\modificaciondos{Although approaches already exist proposing hierarchical clustering for centralised ANN search, such as HCNNG~\cite{vargas2019hcnng}; targeting scalable vector search via HNSW-based local indices, such as Pyramid~\cite{deng2019pyramid}; and even combining both characteristics through hierarchical $k$-means trees and a distributed ANN search framework, such as FLANN~\cite{Muja2014}, they share a common limitation: they are designed for conventional metric or vector-space settings and do not explicitly target arbitrary dissimilarity functions.}

Thus, while further investigation into the effect of different distance functions on ANN search is well motivated, there is a need for flexible methods that can accommodate them. In the following section, we build upon these insights to introduce our approach, which directly addresses these challenges.

\section{THE PROPOSED ALGORITHM}
\label{sec:proposed_algorithm}
In this work, we present PDASC, a fully distributable ANN search algorithm capable of operating under a broad range of distance functions. Our approach relies on applying conventional clustering techniques in a tailored manner to construct a hierarchical index specifically designed to address the challenges inherent to high-dimensional spaces.

\begin{algorithm}[htb]\scriptsize
\caption{Multilevel Structure Algorithm}
\label{algo:MSA}
\begin{algorithmic}[1]
\algnewcommand{\LineComment}[1]{\State \(\triangleright\) #1}
                     
\Procedure{MSA}{$nNodes, dataset, gl, np, distFunc, clusterAlg$}
\LineComment{Split dataset into $nNodes$ size-balanced parts}
\State $dataList \gets \Call{split}{dataset, nNodes}$
\State $nodeList \gets [\ ]$
\For{$node=1, nNodes$}
    \State $data \gets dataList[node]$
    \State $index \gets$ \Call{CreateIndex}{$data, gl, np, distFunc, clusterAlg$}
    \State $nodeList.add(index)$
\EndFor
\State \Return{$nodeList$}
\EndProcedure
                    
\Procedure{CreateIndex}{$data, gl, np, distFunc, clusterAlg$}
\LineComment{Divide dataset into $nGroups$ groups of $gl$ points}
\State $vector, nGroups \gets$ \Call{split}{$data, gl$}
\LineComment{Initialise multilevel structure}
\State $levelPoints \gets [\ ]$
\State $idLevel\gets 0$
                
\While{$nGroups \geq 1$}
    \LineComment{Initialise auxiliary structures for each level} 
    \State $groupsPoints, points \gets [\ ]$
            
    \For{$idGroup=1, nGroups$}
        \State $points \gets vector[idGroup]$
        \State $groupsPoints.add($
        \State \hspace{0.2cm} $clustering($
        $clusterAlg, distFunc, np, points$))                       
    \EndFor
    \State $levelPoints.add(groupsPoints)$
    \State $vector, nGroups \gets$ \Call{split}{$levelPoints, gl$}                  \State $idLevel \mathrel{+}= 1$
\EndWhile
\State $nLevels \gets idLevel-1$
\State \Return{(levelPoints, nLevels)}
\EndProcedure
\end{algorithmic}
\end{algorithm}

\modificacion{The manifestations of the \textit{curse of dimensionality}, particularly distance concentration and the progressive loss of global contrast, reduce the discriminative power of global similarity relationships in high-dimensional spaces. Nevertheless, local neighbourhood relationships, while remaining sufficiently informative for approximate retrieval tasks~\cite{saul2003}, are generally less affected by these phenomena~\cite{houle2010}, making them a natural foundation for designing indexing structures in such settings.

To this end, PDASC constructs a hierarchical clustering-based index in which the data is progressively organised into groups across multiple levels of abstraction. A key design choice is the use of a relatively large set of prototypes to represent each cluster (typically 20--30\% of the items in that group), leading to more stable and well-defined clusters~\cite{powell2022}. This denser representation preserves richer local neighbourhood information and improves the separability of data regions across the hierarchy, facilitating more reliable pruning decisions during search traversal. As a result, the index is expected to be less sensitive to distance concentration, enabling more reliable candidate retrieval and more robust behaviour in high-dimensional settings.}

Furthermore, PDASC can operate directly on heterogeneous data by leveraging clustering methods compatible with non-metric and domain-specific dissimilarity measures. This flexibility makes the proposed method particularly useful in diverse scenarios where the most appropriate similarity measure is not readily determined.

\subsection{PDASC}
\label{subsec:PDASC}


At its core, PDASC operates in two phases: the Multilevel Structure Algorithm (MSA) for building the index, and the Neighbours Search Algorithm (NSA) for processing queries.



As outlined in Algorithm~\ref{algo:MSA}, MSA follows a bottom-up strategy to construct a multilevel index structure tailored for distributed, scalable data indexing. The process begins by partitioning the dataset into size-balanced subsets, which are then distributed across the available computational nodes. To build the lower level of the local index on each \texttt{nNodes} computing node,  the assigned subset is randomly divided into groups of size \texttt{gl}. Each group is clustered independently, and for each resulting cluster, a set of representative prototypes is generated, effectively summarising the local data distribution. These prototypes constitute the first level of the local index on that node.

After completing this level, each set of representative prototypes is assembled with as many consecutive groups as needed to form a new group whose size is as close to \texttt{gl} as possible. These new groups are clustered again to construct the next upper level of the hierarchy and this procedure is recursively applied until reaching a layer where the number of prototypes equals \texttt{np}. Once this process concludes, each computing node maintains an independent multilevel index, and the collection of all local structures constitutes the fully distributed index.
The structure generated by MSA depends on several index parameters:

\begin{itemize}
    \vspace{-0.6em}
    \item \textbf{Distance function} \texttt{(distFunc)}. Specifies the dissimilarity measure used to compute pairwise distances during index construction and search. The set of supported distances is determined by \texttt{clusterAlg}, whose optimisation routines determine which dissimilarities can be handled efficiently.

    \item \textbf{Clustering algorithm} \texttt{(clusterAlg)}.
    Defines the grouping and prototype-selection strategy used during index construction. \modificacion{The chosen algorithm must satisfy three requirements: support for a broad range of dissimilarity measures; generation of prototype-based representations in which each representative is a real dataset element, rather than a synthetic approximation; and explicit control over the number of prototypes produced per group. In this work, $k$-medoids is employed, as it naturally satisfies all three requirements, being $k$ the number of prototypes generated per group \texttt{(np)}.}
    
    \item \textbf{Group length} \texttt{(gl)}. Specifies the number of points in each group over which \modificacion{\texttt{clusterAlg} operates during index construction.}
    
    \item \textbf{Number of prototypes per group} \texttt{(np)}. Determines the number of clusters that \modificacion{\texttt{clusterAlg}} must produce within each group of size \texttt{gl}.

    \item \textbf{Number of computing nodes} \texttt{(nNodes)}. Defines the number of nodes over which the architecture is distributed.

\end{itemize}

\begin{figure}[htb]
    \centering
    \includegraphics[width=0.9\columnwidth]{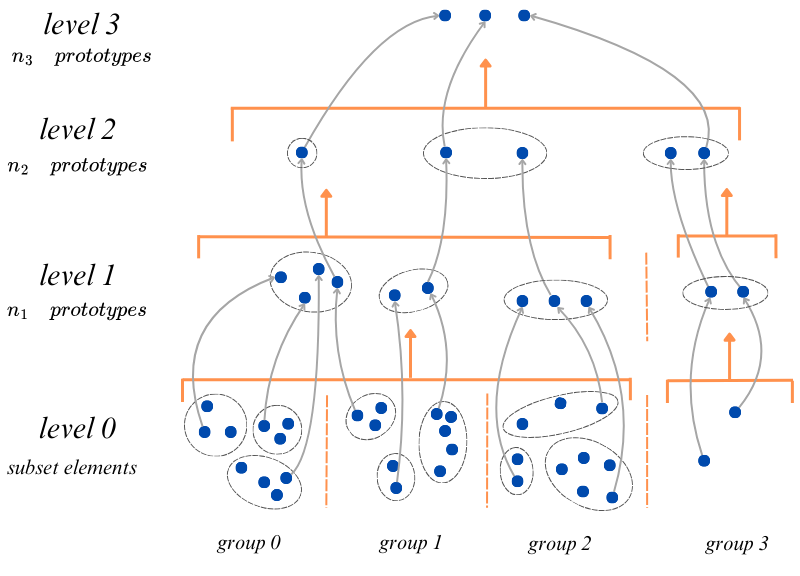}
    \caption{Conceptual representation of the multilevel index construction in PDASC using the MSA.}
    \label{figure-multilayer}
\end{figure}

Fig. \ref{figure-multilayer} illustrates the index-building process of the MSA algorithm. In this example, a dataset of 320 elements is distributed across \texttt{nNodes} = 10 computing nodes, resulting in ten subsets of 32 elements each. For clarity, the figure only depicts the index construction performed on a single computing node.

The parameters are set to a ratio \texttt{np/gl} = 3/10. At the lowest level, the 32-element subset is partitioned into groups of size \texttt{gl}, producing three groups of 10 elements and a residual group of 2 elements. Each group is then clustered into \texttt{np} clusters (indicated by dashed circles), and one prototype is selected from each cluster and promoted to the next level.

At level 1, the prototypes produced by groups 0, 1, and 2 at level 0 are merged to form the first group of that layer. Generally, the algorithm combines the prototypes of a group with those of as many consecutive groups as needed to form a new group that is as close in size to \texttt{gl} as possible without exceeding its value. This procedure is applied recursively, producing progressively more compact prototype layers that summarise the level immediately below. The process concludes when a level containing exactly \texttt{np} points is formed (level 3 in this example).

The algorithm naturally handles variability in group sizes, a phenomenon particularly common in the lowest layers. When a group contains fewer than \texttt{np} elements, all of them are promoted. In this example, group 3 at level 0 contains only two elements, so both are promoted to level 1. Consequently, the group formed by the prototypes of groups 2 and 3 contains only five points, from which three prototypes are selected and promoted to the next level.

This procedure is executed independently on each of the \texttt{nNodes}, yielding a fully distributed hierarchical index.

\begin{algorithm}[htb]
\scriptsize
\caption{Neighbours Search Algorithm (NSA)}
\label{algo:NSA}
\begin{algorithmic}[1]
\Procedure{NSA}{$nodeList$, $distFunc$, $q$, $k$, $r$}
  \State $Candidates \gets \emptyset$
    \For{each $node$ in $nodeList$}
      \State $idLevel \gets getTopLevel(node)$
      \State $prototypes \gets getPrototypes(node, idLevel)$
      \State $D \gets$ distance($q$, $prototypes$, $distFunc$)
      \State $filteredPrototypes \gets$ $D \leq r$
      \State $nodeCandidates \gets$ \Call {ExploreCandidates}{$filteredPrototypes$, $distFunc$, $q$, $r$, $idLevel-1$}
      \State $D \gets$ distance($q$, $nodeCandidates$, $distFunc$)
      \State $Candid. \gets Candid. \cup \{(c, d) \;|\; c \in nodeCandidates, d \in D\}$
    \EndFor
  \State $sortedCandidates \gets$ sort($Candidates$)
  \State $neighbours \gets sortedCandidates[0:k]$
  \State \Return{$neighbours$}
\EndProcedure
  
\Procedure{ExploreCandidates}{$candidates$, $distFunc$, $q$, $r$, $idLevel$}
  \If{$idLevel = 0$}
    \State \textbf{return} $candidates$
  \EndIf
  \State $newCandidates \gets \emptyset$
  \For{each $candidate$ in $candidates$}
    \State $mappedPoints \gets getPrototypes(candidate, idLevel)$
    \State $D \gets$ distance($q$, $mappedPoints$, $distFunc$)
    \State $filteredCandidates \gets$ $D \leq r$
    \If{$filteredCandidates \neq \emptyset$}
      \State $result \gets$ \Call{ExploreCandidates}{$filteredCandidates$, $distFunc$, $q$, $r$, $idLevel-1$}
      \State $newCandidates \gets newCandidates \cup \{result\}$ 
    \EndIf
  \EndFor
\State \Return{$newCandidates$}
\EndProcedure
\end{algorithmic}
\end{algorithm}

\begin{figure*}[!htb]
    \centering
    \includegraphics[width=0.85\textwidth]{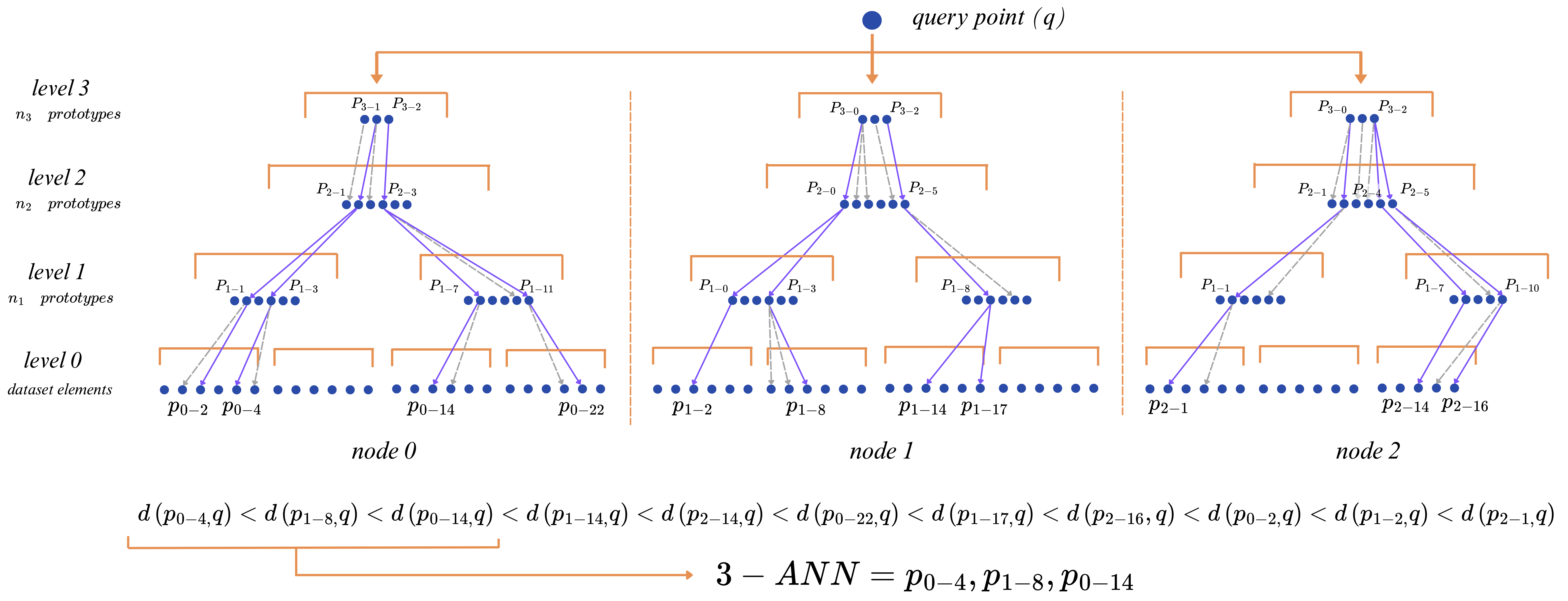}
    \caption{Conceptual illustration of the $k$-NN search in PDASC using the NSA over a multilevel index distributed across $nNodes = 3$ computing nodes. The dataset of 65 elements is partitioned into subsets of 24 (\textit{node 0}), 24 (\textit{node 1}), and 11 elements (\textit{node 2}), where each local index is built using $gl = 6$ and $np = 3$.}
    \label{figure-search-example}
\end{figure*}

The NSA algorithm performs a top-down traversal of the multilevel index generated by the MSA on each computing node to efficiently retrieve candidates for the $k$-ANN of a query point $q$, as described in Algorithm~\ref{algo:NSA}. At each level, for every prototype selected for exploration, distances between $q$ and its child prototypes (i.e. those associated with that prototype at the next lower level) are computed. Only the child prototypes whose distance to $q$ is less than or equal to a predefined global radius $r$ are selected for further exploration. This traversal continues recursively, progressively narrowing the search space until the bottom level is reached. At this lowest level, the corresponding data points are collected into the candidate pool. Once all computing nodes have completed their local traversals and calculated the distances from $q$ to all candidates, such candidate sets are combined into a single pool. The final pool is then sorted to identify the top-$k$ ANN (see Fig.~\ref{figure-search-example}).

\modificacion{Thus, the behaviour of the NSA algorithm depends on the \textbf{search radius} \texttt{(r)} parameter, which defines the distance threshold used to decide whether a prototype should be explored during index traversal. The value of $r$ can be determined from statistics derived from the dataset distance distribution, such as the cumulative distribution function or the maximum observed distance.}

Fig. \ref{figure-search-example} illustrates an example of the search process for retrieving the $3-ANN$ of $q$. The global PDASC index is distributed across three computing nodes, each storing the portion of the index built from its local data partition. During query processing, NSA is executed independently on the index stored at each computing node. In the Figure, $P_{x-y}$ denotes the $y$-th prototype at level $x$ of the index located in a computing node, and $p_{z-w}$ denotes the $z$-th data point in the subset located on computing node $w$. For clarity, the explanation focuses on the search performed on node 0; while the same behaviour generalises to all other nodes.

The search process begins at the highest level of the index (level 3). NSA computes the distances between the query point $q$ and all prototypes at this level (that is, $d(P_{3-0}, q)$, $d(P_{3-1}, q)$, etc.) and retains only those satisfying $d(q, P_{3-i}) \leq r$ for further exploration. In Fig. \ref{figure-search-example}, prototypes that satisfy this condition and give rise to valid descent paths are represented by solid lines, whereas branches pruned due to distances exceeding the threshold are depicted with dotted lines. In the illustrated example, the prototypes $P_{3-1}$ and $P_{3-2}$ meet the threshold and are therefore explored at the next level.

The algorithm then evaluates the child prototypes mapped by each selected prototype, retaining only those that satisfy the distance condition. For example, $P_{3-1}$ maps to \textit{level 2} prototypes $(P_{2-1}, P_{2-2})$, but only $P_{2-1}$ satisfies $d(q, P_{2-1}) \leq r$. The same behaviour occurs for the prototypes mapped by $P_{3-2}$.

Then, the algorithm continues recursively, expanding only the prototypes that meet the distance constraint. In this example, the valid descent paths from $P_{2-1}$ and $P_{2-3}$ lead to \textit{level 1} prototypes $P_{1-1}$, $P_{1-3}$, $P_{1-7}$, and $P_{1-11}$, from which the search eventually reaches the corresponding points at \textit{level 0}. These points constitute the candidate set retrieved on \textit{node 0}, which is then merged with the candidates obtained from \textit{node 1} and \textit{node 2}.

After merging, all candidate points are globally sorted by their distance to $q$, and the three closest points are selected as the final nearest neighbours:  $\textrm{3-ANN}=\{p_{0-4},\, p_{1-8},\, p_{0-14}\}$

\subsection{\modificacion{FACTORS GOVERNING PDASC BEHAVIOUR}}
\label{subsec:index_behaviour}
The behaviour of the PDASC index is governed by the parameters introduced in Section~\ref{subsec:PDASC} and, crucially, by the interactions among some of them. These factors determine the topology of the hierarchy, the neighbourhood relationships captured by the index, and the traversal dynamics during search, ultimately influencing recall, NDC and index size. Beyond the qualitative insights discussed below, the observed effects are supported by the experimental results reported in Section~\ref{sec:experimental_evaluation}.

\textbf{Number of computing nodes} \texttt{(nNodes)}.
This parameter controls the level of parallelism and data partitioning. Increasing \texttt{nNodes} produces smaller local subsets, reducing clustering and search workloads per node, potentially improving computational efficiency during search. However, these gains exhibit diminishing returns, since excessive partitioning progressively fragments the data space and limits the effectiveness of the hierarchy. 

\textbf{Group length} \texttt{(gl)}. This parameter determines the number of points processed during each local clustering step, directly affecting the computational cost of index construction.  Small values generate many small groups, increasing clustering overhead, whereas very large values increase the internal cost of clustering within each group. 
Experimental results across datasets with different densities confirm that the most suitable values are data-dependent.

\textbf{Number of prototypes per group} \texttt{(np)}.This parameter determines the amount of information each group contributes to the upper levels, hence defining the structural granularity of the resulting index. Larger values of \texttt{np} increase index size and indexing time by propagating more representatives. 

\textbf{Prototype-to-group ratio} \texttt{(np/gl)}.
Although the isolated influence of \texttt{gl} and \texttt{np} on recall and NDC is limited, their interaction determines the prototype density propagated through the hierarchy, which ultimately governs the resulting index topology and its recall-NDC trade-off. Low ratios produce compact hierarchies and increase the proportion of the index explored during search, yielding high recall at the cost of higher NDC (indeed, extremely low ratios approach full-scan behaviour). Conversely, high ratios propagate too many representatives to upper layers, leading to excessive partitioning and less faithful cluster representations. As a consequence, pruning decisions become overly restrictive, since the upper-level structure no longer reflects the underlying data distribution accurately, ultimately reducing recall. In addition, the index size increases with the ratio as more prototypes are propagated upward through the hierarchy. Therefore, effective index construction requires a balanced ratio that preserves sufficient search-space coverage without incurring over-partitioning.

\modificacion{\textbf{Distance function} \texttt{(distFunc)} and \textbf{clustering algorithm} \texttt{(clusterAlg)}. Beyond the topological parameters discussed above, the clustering method and the selected distance function directly influence prototype selection, shaping the geometric structure of the resulting partitions and ultimately determining the neighbourhood relationships captured by the index.}

\modificacion{\textbf{Search radius} \texttt{(r)}.The radius parameter governs the restrictiveness of the hierarchical traversal during search, directly controlling the recall-NDC trade-off. Small values of \texttt{r} aggressively prune the search space, reducing computational cost at the expense of potentially missing relevant neighbours. In contrast, larger values of \texttt{r} explore broader regions of the hierarchy, generally improving recall while requiring a higher number of distance computations.}

\begin{table}[h]
\centering
\caption{\modificacion{Summary of the main factors governing PDASC index behaviour.}}
\label{tab:pdasc_parameter_effects}

\resizebox{\columnwidth}{!}{%
\begin{tabular}{l l l l}
\toprule
\textbf{Parameter} & \textbf{Category} & \textbf{Primary Effect} & \textbf{Impact on Index Behaviour} \\
\midrule

\texttt{nNodes} & Structural &
Data partitioning &
Scalability and per-node index size \\

\texttt{gl} & Structural &
Local granularity &
Clustering cost and locality \\

\texttt{np} & Structural &
Representation density &
Index size and representational capacity \\

\texttt{np/gl} & Derived &
\textbf{Topology control} &
Coverage vs. over-partitioning (Recall--NDC trade-off) \\

\texttt{distFunc} & Algorithmic &
Similarity definition &
Neighbourhood structure and prototype selection \\

\texttt{clusterAlg} & Algorithmic &
Prototype selection &
Prototype fidelity and partition quality \\

\texttt{r} & Search &
Traversal restrictiveness &
Exploration breadth (Recall--NDC trade-off) \\

\bottomrule
\end{tabular}%
}
\end{table}

\subsection{\modificacion{COMPLEXITY ANALYSIS}}
\label{subsec:complexity-analysis}

 \modificacion{
To analyse the asymptotic complexity of PDASC, we consider the time complexity of index construction, the time complexity of approximate nearest neighbour search, and the space complexity of the distributed multilevel index. Here, $n$ denotes the number of indexed vectors and $d$ the vector dimensionality.

The hierarchical index construction is performed independently at each computational node. Each local index recursively partitions its assigned data into groups of size $gl$ and promotes $np$ representative prototypes per group to the next level. Assuming a FasterPAM $k$-medoids clustering cost of $O(gl^2 \cdot d)$ per group, the number of points decreases geometrically by a factor $np/gl$ at each level, yielding the overall parallel construction complexity

\[
O\!\left(
\frac{n}{nNodes}\cdot \frac{
gl  \cdot d
}
{
1-\frac{np}{gl}
}
\right).
\]

For fixed $gl$, $np$, and $nNodes$, the complexity reduces to $O(n \cdot d)$.

Regarding query processing, the search complexity depends on the effectiveness of the radius-based pruning strategy. Assuming that, on average, a fraction $\alpha$ of the candidates survives the pruning process at each level, where $\alpha$ is determined by the selected search radius $r$, the search complexity is

\[
O\!\left(
np \cdot d
\sum_{h=0}^{L}
\left(
\alpha\frac{gl}{np}
\right)^h
+
\kappa \cdot nNodes \cdot \log(\kappa \cdot nNodes)
\right),
\]
where the first term corresponds to the parallel traversal of the distributed hierarchy, whereas the second term accounts for merging the $\kappa$ nearest neighbours returned by each computational node. Here, $L$ denotes the number of levels in the local hierarchy, approximately given by

\[
L \approx
\log_{gl/np}
\!\left(
\frac{n}{nNodes\cdot np}
\right).
\]





If $gl$, $np$, $k$, and $nNodes$ are treated as fixed configuration parameters, the search complexity becomes independent of the dataset size under effective pruning ($\alpha gl/np < 1$), reducing to $O(d)$. In the worst case ($\alpha = 1$), the complexity degrades to $O(n \cdot d)$.
}

\modificacion{Finally, the space complexity of PDASC is determined by the storage of the multilevel hierarchy. Since each level reduces the number of stored elements by a factor $np/gl$, the resulting memory requirements form a geometric series. Assuming an even distribution of the data across $nNodes$ computational nodes, the space complexity per node is

\[
O\!\left(
\frac{n}{nNodes}
\cdot
\frac{d}
{1-\frac{np}{gl}}
\right).
\]



If $gl$, $np$, and $nNodes$ are treated as fixed configuration parameters, the space complexity becomes linear with respect to the number of indexed vectors, i.e., $O(n \cdot d)$.}

 \modificacion{
Table~\ref{tab:ann_complexities} compares the theoretical complexities of PDASC and the ANN methods considered in this study. For LSH, $L$ and $K$ denote the number of hash tables and hash functions per table, respectively. For IVF, $k$ and $P$ represent the number of centroids (\textit{nlist}) and probed inverted lists (\textit{nprobe}). For HNSW, $M$ denotes the maximum number of graph connections per node, while $t$ represent the number of trees in ANNOY. Finally, for PyNNDescent, $q$ denotes the number of stored neighbours per point and $\alpha > 1$ its empirical scaling exponent.
}

\begin{table}[ht]
\centering
\caption{\modificacion{Comparison of the theoretical complexities of the ANN algorithms considered. }}
\label{tab:ann_complexities}
\resizebox{\columnwidth}{!}{%
\begin{tabular}{llll}
\hline
\textbf{Algorithm} & \textbf{Build Complexity} & \textbf{Search Complexity} & \textbf{Space Complexity} \\
\hline

ANNOY &
$O(n \cdot t \cdot \log n \cdot d)$ &
$O(t \cdot \log n \cdot d)$ &
$O(n \cdot d \cdot t)$ \\

HNSW &
$O(n \cdot \log n \cdot d)$ &
$O(\log n \cdot d)$ &
$O(n \cdot (d + M))$ \\

IVF &
$O(n \cdot k \cdot d)$ &
$O\!\left(k \cdot d + P \cdot \frac{n}{k} \cdot d\right)$ &
$O((n + k) \cdot d)$ \\

LSH &
$O(n \cdot L \cdot K \cdot d)$ &
$O(n^{\rho})$ &
$O(n \cdot (d + L))$ \\

PDASC &
$O(n \cdot d)$ &
$O(d)$ (best); $O(n \cdot d)$ (worst) &
$O(n \cdot d)$ \\

PyNNDescent &
$O(n^{\alpha} \cdot d)$ &
$O(\log n \cdot d)$ &
$O(n \cdot (d + q))$ \\
\hline
\end{tabular}
}
\end{table}

\modificacion{As observed, PDASC provides favourable construction and memory scalability compared to several graph- and tree-based ANN methods, while its search performance depends on the effectiveness of radius-based pruning. Under effective pruning, the search cost can become independent of the dataset size, whereas in the worst case it degenerates to a linear scan. Therefore, PDASC's main strength does not lie in a uniformly superior theoretical search bound, but rather in its balanced trade-off between memory efficiency, distributed construction, and distance-function flexibility.}

\section{EXPERIMENTAL STUDY}
\label{sec:experimental_evaluation}

This section presents a comprehensive empirical evaluation of PDASC. Experiments are conducted on the datasets listed in Table~\ref{tab:dataset_features}, selecting distance functions that reflect the semantic notion of similarity inherent to each dataset while assessing PDASC’s support for arbitrary dissimilarity measures.

The evaluation is organised into two parts. First, we analyse how different parametrisations shape the hierarchical index structure, influencing its topology, size, and the behaviour of the search algorithm. Second, we compare PDASC against state-of-the-art ANN algorithms to contextualise its performance, in terms of both computational cost and memory footprint.

For quality assessment, we report the recall achieved under different parametrisations, measured against the ground-truth set of 10-NN obtained via exact search. For performance assessment, we evaluate both the NDC per computing node and the index size stored on each node.

As distance evaluations constitute the dominant component of query-time cost in most ANN algorithms~\cite{GaoLong_2023,peng2023}, NDC is adopted as the primary performance metric. Moreover, it helps isolate the intrinsic behaviour of the algorithm by reducing the influence of platform-dependent optimisations (such as low-level SIMD vectorisation, cache-aware memory layouts, or multi-threading strategies) that may otherwise bias comparisons, even within the same architecture, thereby providing a clearer view of the underlying computational cost\modificacion{~\cite{kriegel2017}}.

Similarly, index size has been included as a complementary metric, as memory footprint is a critical limiting factor in large-scale ANN systems and becomes even more relevant in low-resource and distributed environments. In such scenarios, index size determines RAM usage, persistent storage requirements, and even the number of machines needed to host the system, directly impacting operational and energy costs. Considering index size, therefore, provides a realistic assessment of scalability and practical viability in distributed infrastructures.

Thus, we will examine how NDC and index size per computing node vary across different parameter settings, presenting recall–NDC and recall–index-size trade-off plots that characterise the relationship between accuracy, computational effort, and memory footprint~\cite{aumuller2021,yang2025}.

\modificacion{\textbf{Implementation and Experimental Environment.}\\
The PDASC algorithm was fully implemented in Python, and all source code related to both the proposed method and the experimental evaluation is publicly available in the }{\href{https://github.com/elenagarciamorato/PDASC}{elenagarciamorato/PDASC} GitHub repository.
For the clustering stage, PDASC relies on the \texttt{FasterPAM} Python implementation~\citep{schubert_fast_2022}, selected for its computational efficiency and support for a broad variety of distance measures, ensuring compatibility with the diverse dissimilarity functions considered in the experiments.}


All experiments were conducted on a server equipped with an AMD EPYC Genoa processor (32 cores / 64 threads), 384 GB of RAM, and 5.76 TB of SSD storage.


\subsection{DATASETS}
\label{subsec:datasets}
Table~\ref{tab:dataset_features} lists the five publicly available datasets used in our experiments, which differ in their type, number of elements ($n$), and dimensionality ($D$). Four of these datasets (MNIST, NYTimes, GLOVE, and MovieLens-10M) are standard benchmarks commonly used to evaluate ANN search implementations~\cite{Aumuller2020}, ensuring comparability with prior work. The \textit{Municipalities} dataset is a two-dimensional geospatial collection comprising the latitude and longitude coordinates of all 8,130 municipalities in Spain\footnote{https://ign.es/web/rcc-nomenclator-nacional}. Hence, this is a representative test case for geospatial queries, allowing the use of the Haversine distance function~\cite{sinnott1984} to further enrich the diversity of distance measures considered in the experimental study. These datasets also exhibit various levels of local intrinsic dimensionality (LID)~\cite{houle2013dimensionality}, confirming the variety and difficulty of our evaluation scenarios~\cite{aumuller2021} and are processed using different distance functions to assess PDASC's support for arbitrary distances, including common metric distances (e.g., Euclidean), less frequently used ones (e.g., Jaccard), and non-metric distances (e.g., cosine).

\begin{table}[htb]
\centering
\caption{Summary of key features of selected datasets.}
\label{tab:dataset_features}
\resizebox{\columnwidth}{!}{%
\begin{tabular}{l r r l c l}
\toprule
\textbf{Dataset} & \textbf{\textit{n}} & \textbf{\textit{D}} & \textbf{Data Type} & \textbf{LID} & \textbf{Distance} \\
\midrule
\textit{Municipalities} & 8,130 & 2 & Geospatial & 2.65 & Haversine \\
MNIST & 69,000 & 784 & Image & 13.05 & Euclidean \\
GLOVE & $10^6$ & 100 & Text & 17.62 & Cosine \\
NYTimes & 290,000 & 256 & Text & 46.88 & Cosine \\
MovieLens10M & 65,134 & 69,363 & Boolean & 22.41 & Jaccard \\
\bottomrule
\end{tabular}%
}
\end{table}

\subsection{STRUCTURAL SENSITIVITY ANALYSIS OF PDASC} 
\label{subsec:experimental-sensitivity_an}

For each dataset, the data is randomly split into non-overlapping training and test subsets to ensure statistical robustness and avoid bias. The training sets are then partitioned and distributed across $nNodes$ computational nodes. In this study, we evaluate configurations with $nNodes \in \{1, 3, 5, 10\}$ to illustrate different levels of parallelism and demonstrate PDASC’s applicability in distributed environments. At each computational node, an index tree is constructed by applying the MSA algorithm to the corresponding subset, using the selected clustering algorithm, distance function, group length (\texttt{gl}), and number of prototype points (\texttt{np}). To assess index structures at different levels of granularity, we evaluate several \texttt{np/gl} ratios, specifically 0.02, 0.05, 0.1, 0.2, 0.33, and 0.5.

Subsequently, the 10-ANN search for each query point ($q$) in the test set is performed using NSA. For each experiment, different values of \texttt{r} are evaluated to explore the behaviour of PDASC under varying levels of search restrictiveness. \modificacion{Since the radius parameter can be selected from statistical characterisations of the dataset distance distribution, in this study we derive representative values from lower and upper percentiles of the empirical cumulative distribution function (ECDF) of pairwise distances. Specifically, $P_{ECDF}=30$ and $P_{ECDF}=70$ are used to represent more and less restrictive search configurations, corresponding to $r=0.91$ and $r=1.04$ for the GLOVE dataset.}

\begin{figure}[h!]
    \centering
    \begin{subfigure}[b]{0.48\columnwidth}
        \centering
        \includegraphics[width=0.9\columnwidth]{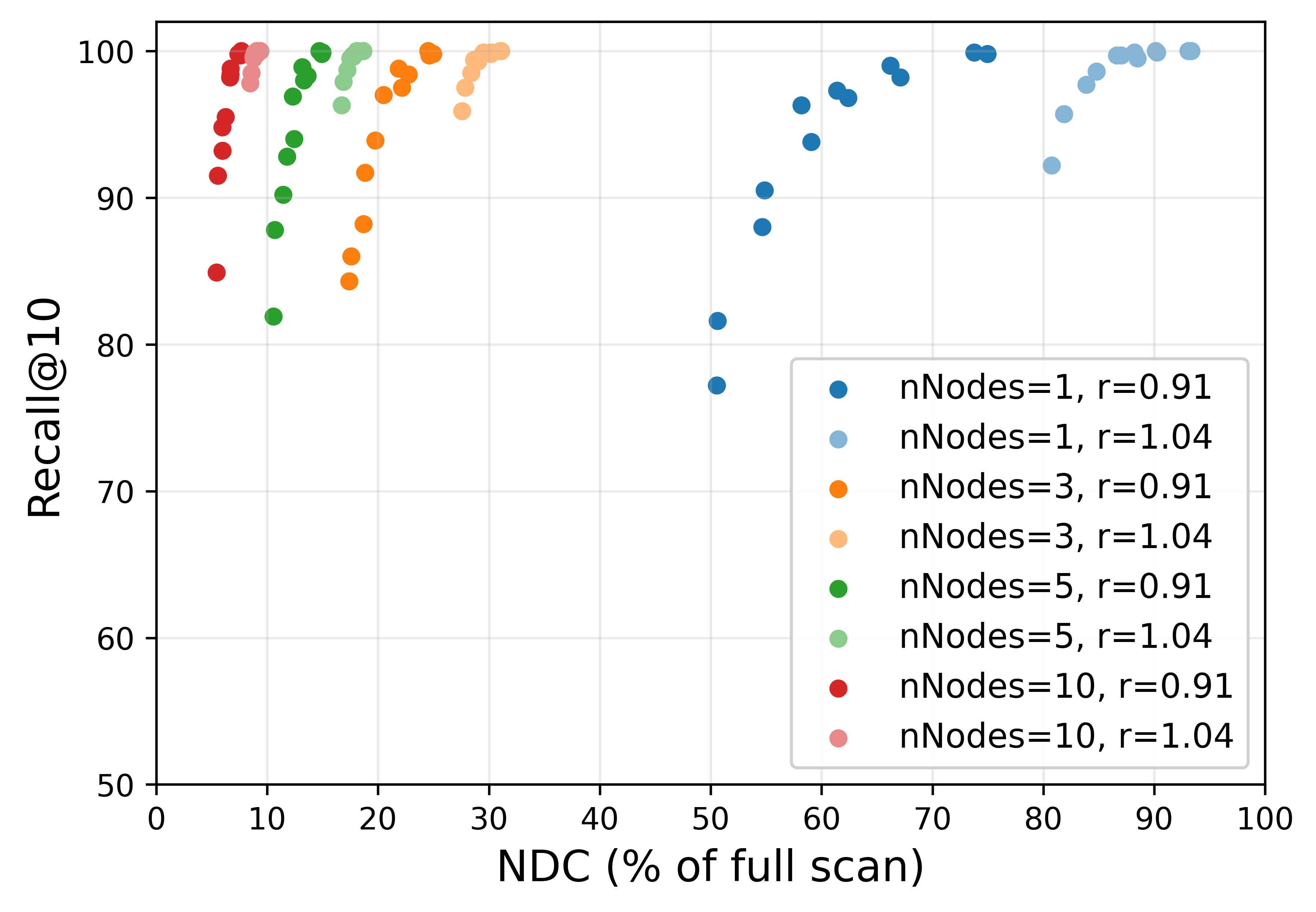}
                \vspace{-0.5em}
        \caption{Recall vs. NDC for different $r$ configurations.}
        \label{fig:Recall-NDC_GLOVE}
    \end{subfigure}
    \begin{subfigure}[b]{0.48\columnwidth}
        \centering
        \includegraphics[width=0.9\columnwidth]{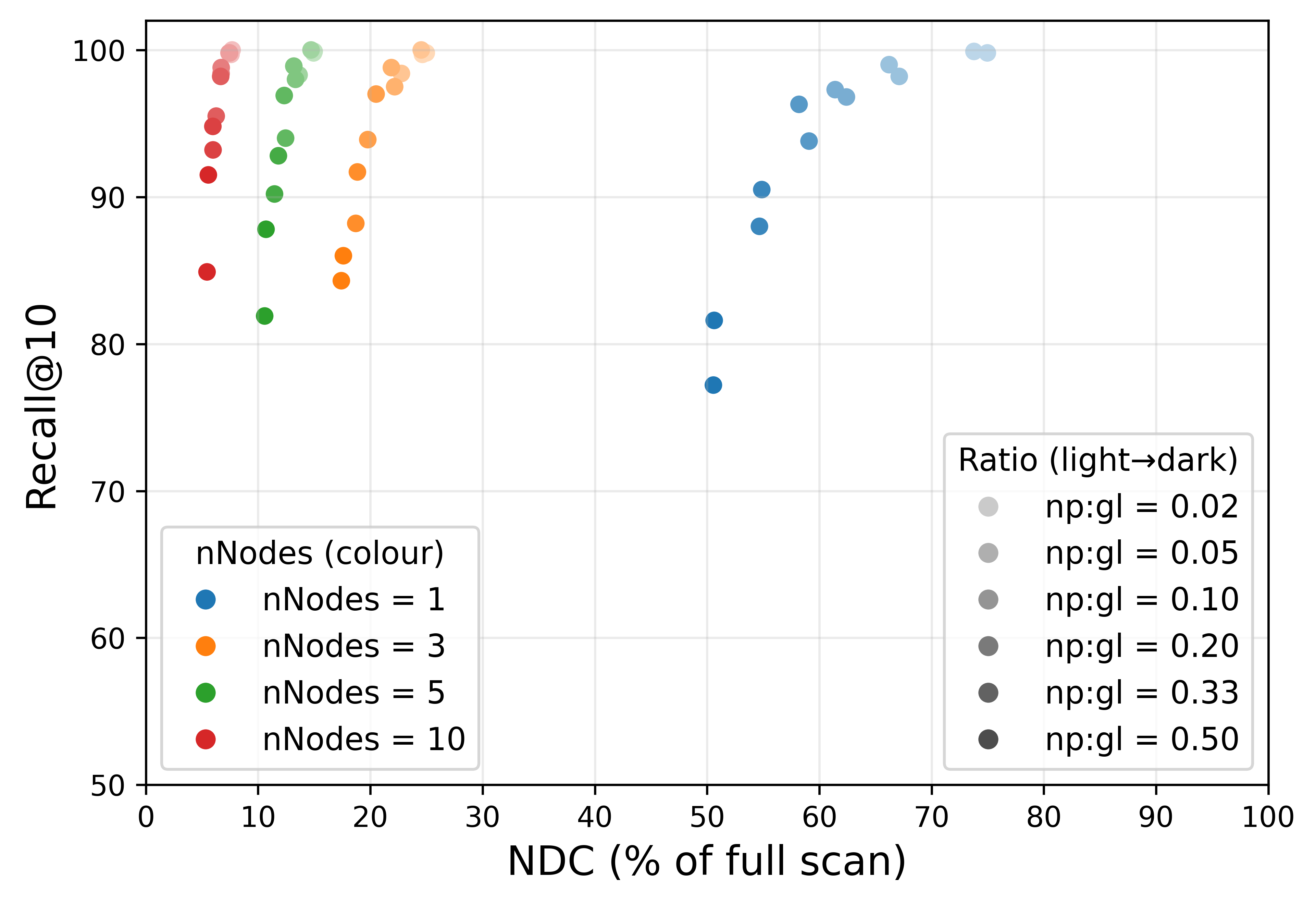}
        \vspace{-0.5em}
        \caption{Recall vs. NDC for different index topologies.}
        \label{fig:Recall-NDC_ratios_GLOVE}
    \end{subfigure}
        \begin{subfigure}[b]{0.5\columnwidth}
        \centering
        \includegraphics[width=0.9\columnwidth]{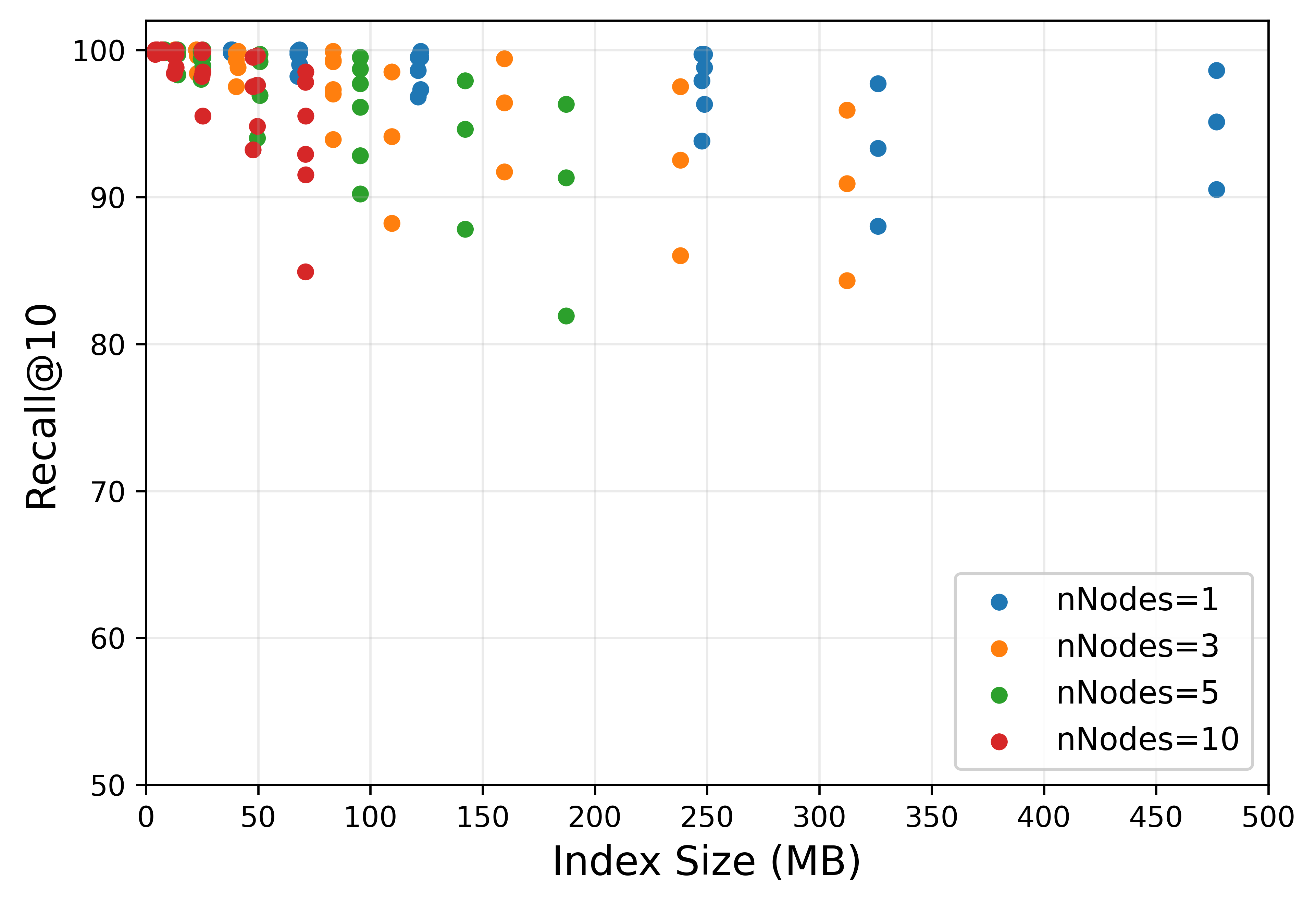}
                \vspace{-0.5em}
        \caption{Recall vs. per-node index size.}
        \label{fig:Recall-IndexSize_GLOVE}
    \end{subfigure}
    \caption{PDASC Performance on GLOVE dataset under different degrees of parallelisation ($nNodes$).}
    \label{fig:Recall_NYtimes}
\end{figure}

Fig.~\ref{fig:Recall-NDC_GLOVE} illustrates how the recall–NDC trade-off per computing node evolves under different PDASC configurations on the GLOVE dataset. Markers sharing the same colour correspond to the same search radius and number of computing nodes. Within each colour group, different data points correspond to different \texttt{np/gl} ratios, thus inducing distinct index topologies. As expected, increasing the number of computing nodes reduces the per-node computational effort, although efficiency gains diminish beyond a certain point. Moreover, larger search radii consistently yield higher recall, generally at the expense of increased NDC. \modificacion{This behaviour is expected, as larger values of \texttt{r} relax the traversal constraints imposed during hierarchical exploration, allowing the search process to access a broader set of candidate regions within the index. However, the impact of \texttt{r} progressively decreases as the number of computing nodes increases, suggesting that finer data partitioning partially compensates for more restrictive search configurations.} Importantly, this chart also reveals that, even with a sufficient number of computing nodes and an appropriately chosen search radius, the overall performance of the algorithm remains strongly dependent on the index configuration, particularly on the choice of the \texttt{np} and \texttt{gl} parameters, whose impact is analysed in more detail in the following figure.

Fig.~\ref{fig:Recall-NDC_ratios_GLOVE} reports the same experimental setting but fixes the search radius to the most restrictive value, $r=0.91$, to isolate and accentuate the influence of the \texttt{np/gl} ratio under stringent search conditions. While it is evident that \texttt{nNodes} and the search radius influence the per-node NDC–recall balance, the \texttt{np/gl} ratio plays a critical role in determining how effectively the hierarchy represents the data, thereby influencing this trade-off. Very large ratios lead to an overly coarse representation, resulting in both low recall and low NDC. As the ratio decreases, prototypes capture local structure more faithfully, yielding a rapid increase in recall at a relatively modest computational cost. Further reductions eventually lead to over-exploration of the hierarchy, as reflected by a steep rise in NDC with only marginal gains in recall. Across all evaluated datasets, intermediate \texttt{np/gl} ratios, typically around 0.2–0.3, consistently provide the most favourable balance among recall, NDC, and index size. These ratios achieve recall values above 90\% while maintaining substantially reduced computational costs and moderate memory requirements. 

Fig.~\ref{fig:Recall-IndexSize_GLOVE} reports the per-node index size of PDASC at each computing node as a function of the number of \texttt{nNodes} employed. While it is evident that the index size is largely governed by the \texttt{nNodes} parameter (increasing \texttt{nNodes} spreads the load and reduces both the per-node and overall memory footprint), certain configurations with fewer nodes achieve comparable or even higher recall (upper-left part of the figure). This is again due to the \texttt{np/gl} ratio, since the index topology not only determines index size but, more importantly, search effectiveness, thereby impacting the trade-off between recall and efficiency.

Overall, the empirical evaluation validates the theoretical characterisation of PDASC presented in Section~\ref{subsec:index_behaviour}: the index topology induced by \texttt{nNodes} and the \texttt{np/gl} ratio governs memory distribution and representativeness, ultimately determining the trade-offs between computational efficiency, pruning behaviour, and search accuracy.

\subsection{BENCHMARKING WITH ALTERNATIVE ANN ALGORITHMS}
\label{subsec:results}
Finally, empirical results are compared against selected state-of-the-art algorithms. 
Therefore, in this study, we include well-established and widely used approaches from different ANN families, all available through Python implementations: ANNOY, PyNNDescent, the FAISS~\cite{douze2025} implementations of LSH and IVF and the Python bindings of the HNSW implementation from nmslib~\cite{MalkovY20}.

\begin{figure*}[htb]
    \centering
    \includegraphics[width=1\linewidth]{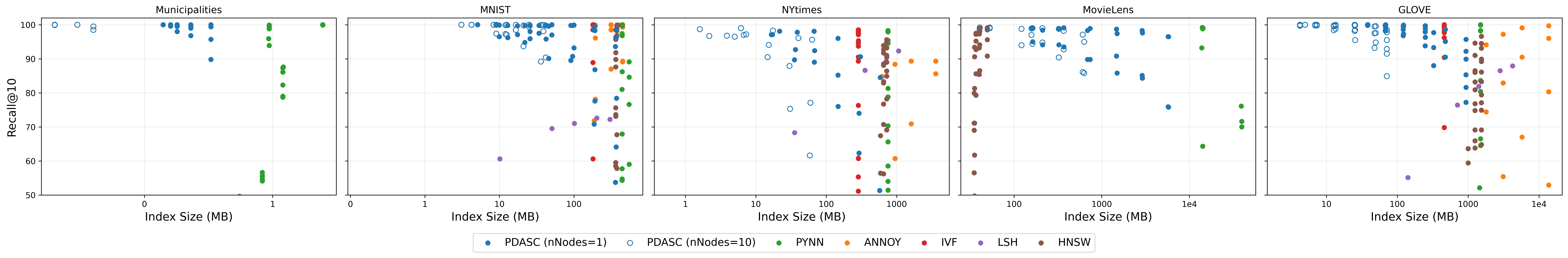}
    \caption{Per-node recall vs. index size for PDASC and competing ANN methods across all datasets (logarithmic x-axis).}
    \label{fig:idx_recall_all_datasets}
\end{figure*}

Fig.~\ref{fig:idx_recall_all_datasets} presents, for all datasets considered in this study, the per-node recall-index size profiles of the selected SOTA methods under different parameter settings, together with the PDASC configurations for \texttt{nNodes}=1 and \texttt{nNodes}=10, which represent the most extreme cases evaluated, ranging from a fully centralised index to a highly distributed deployment. Notably, only PyNNDescent, HNSW, and PDASC are compatible with MovieLens-10M, while only PDASC and PyNNDescent support the \textit{Municipalities} dataset, due to the use of Jaccard and Haversine distances respectively, metrics unsupported by most conventional ANN frameworks. \modificacion{For visual clarity, the y-axis is restricted to recall values above 50, since lower-accuracy configurations are generally of limited practical interest.}

This figure illustrates not only the search quality achieved by different methods but also how efficiently each method uses the available resources at node level, revealing distinct behaviours across algorithms. Methods such as HNSW and IVF exhibit an almost constant per-node index size, whereas their recall varies depending on their internal configuration. Conversely, ANNOY and PyNNDescent exhibit greater variability in memory footprint while attaining a broad range of recall levels regardless of index sizes. LSH shows the greatest variability in memory footprint across parameter settings and is also the method that consistently delivers substantially lower recall. Across all datasets, the PDASC index remains smaller than, or comparable in size to, all evaluated methods at equivalent recall levels, even when the entire PDASC index is concentrated on a single node, thereby highlighting its memory-efficient design.

\modificaciondos{Table~\ref{tab:recall_summary} summarises recall across configurations and experimental
repetitions on NYTimes. Mean and SD denote average recall and variability,
$P_{5}$--$P_{95}$ the central 90\% of observations, and the last column
the percentage of experiments with recall $\geq 90\%$.
PDASC achieves the highest mean recall (86.3\%) and lowest variability (16.5\%), indicating comparatively accurate and consistent approximation performance across the evaluated settings.}
\modificaciondos{Since PDASC is a randomised approximate method, we also analyse how frequently a predefined approximation error threshold is achieved. Defining the approximation error as $1-\mathrm{recall}$, an error of at most 10\% corresponds to a recall greater than or equal to 90\%, a threshold satisfied by PDASC in 61.7\% of the evaluated cases. Results for the remaining datasets are reported in the Supplementary Material, together with a further analysis showing that the approximation quality of PDASC strongly depends on the $np/gl$ ratio, with lower ratios generally associated with higher recall.}

\begin{table}[htb]
\centering
\caption{Summary statistics of recall (\%) for the different algorithms on the NYTimes dataset.}
\label{tab:recall_summary}
\resizebox{\columnwidth}{!}{
\begin{tabular}{lrrrrrrr}
\toprule
Algorithm & Mean & SD & Min & $P_{5}$ & $P_{95}$ & Max & Recall $\geq 90\%$ (\%) \\
\midrule
ANNOY & 57.0 & 25.3 & 19.7 & 22.0 & 97.2 & 99.0 & 19.3 \\
HNSW  & 55.1 & 35.2 &  0.1 &  0.4 & 94.5 & 96.9 & 16.3 \\
IVF   & 85.1 & 19.3 & 40.0 & 45.4 & 99.1 & 99.4 & 67.7 \\
LSH   & 60.1 & 35.6 &  0.0 &  0.2 & 91.9 & 92.7 & 23.8 \\
PDASC & 86.3 & 16.5 & 19.2 & 47.6 & 98.8 & 99.4 & 61.7 \\
PYNN  & 21.7 & 29.9 &  2.5 &  3.7 & 97.7 & 99.4 & 10.0 \\
\bottomrule
\end{tabular}
}
\end{table}

\begin{figure}[htb]
    \centering
    \includegraphics[width=0.85\columnwidth]{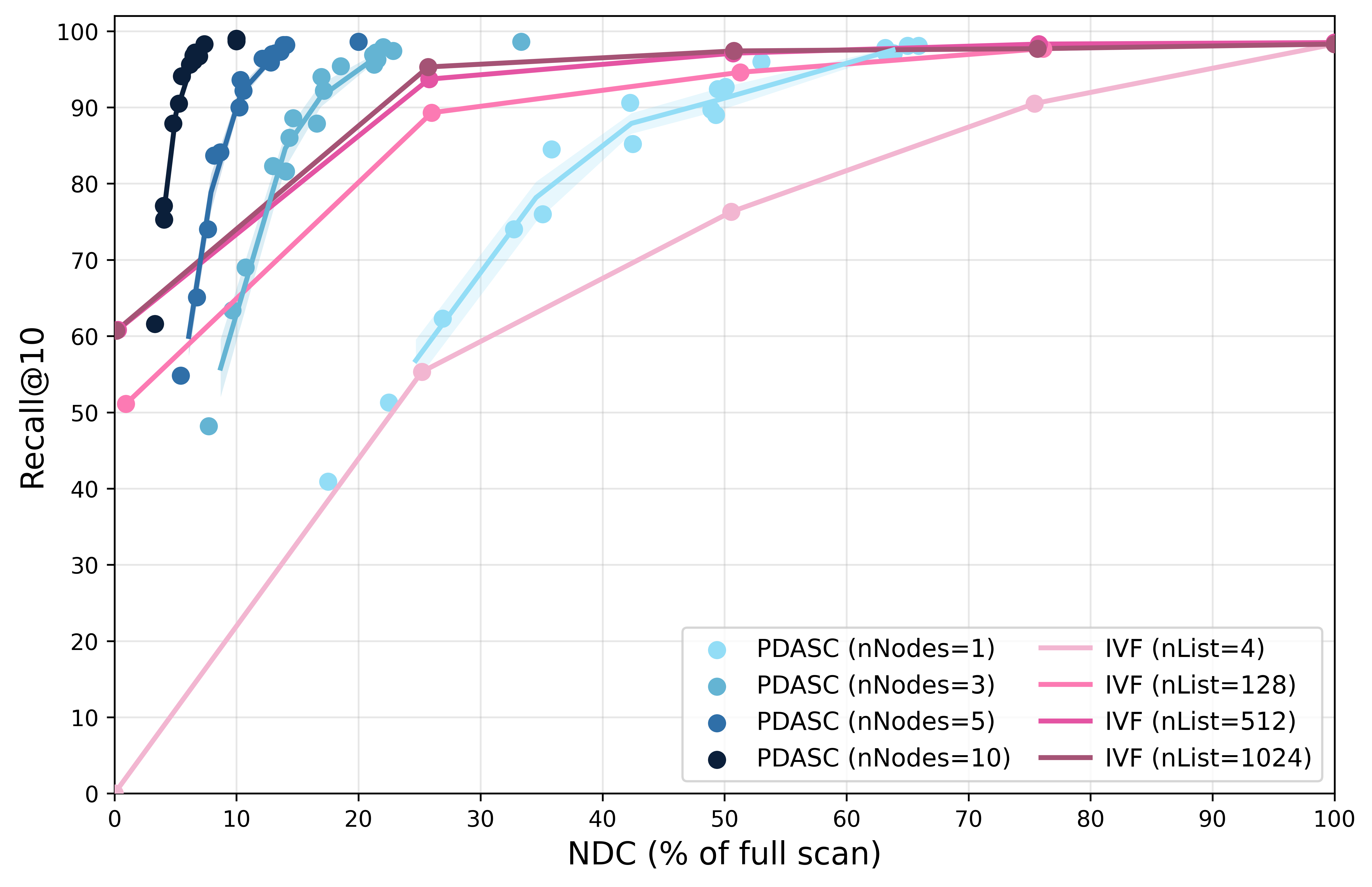}
    \caption{Recall vs. NDC for PDASC and IVF on the NYTimes dataset.}
    \label{figure-recall-NDC-PDASC-IVF}
\end{figure}

\modificacion{Regarding the computational-cost assessment, NDC is not uniformly reported across ANN frameworks: ANNOY, PyNNDescent, HNSW, and LSH do not expose the NDC performed during query execution, preventing an implementation-independent assessment of their intrinsic computational cost. Consequently, the NDC–recall analysis presented in Fig.~\ref{figure-recall-NDC-PDASC-IVF} is restricted to IVF and PDASC, as these are the only evaluated methods that explicitly report NDC as a query-time performance metric.}

For IVF, each line corresponds to a different \texttt{nlist} configuration, while points along each line represent increasing values of \texttt{nprobe}, fixed at 0\%, 25\%, 50\%, 75\%, and 100\% of the selected \texttt{nlist}. For PDASC, each line corresponds to a different \texttt{nNodes} setting \modificacion{under a restrictive search radius configuration corresponding to $P_{ECDF}=30$,} and each point denotes a distinct \texttt{np/gl} ratio, inducing a different index topology.

Therefore, for comparable recall levels, PDASC requires significantly fewer NDC per node than IVF. In PDASC, increasing the number of computing nodes \texttt{nNodes} effectively shifts the operational curves toward the left, demonstrating that as the algorithm better exploits its parallelisation capability, the recall–NDC trade-off improves. Conversely, IVF shows a wider dispersion and higher local computational demand to reach equivalent accuracy.  This comparison highlights PDASC's ability to maintain high search quality while keeping local resource requirements low, making it particularly suitable for environments where per-node computational power is a constraint. 

\modificaciondos{Regarding wall-clock runtime, comparisons with single-node methods should be interpreted with caution, as PDASC follows an embarrassingly parallel execution model designed to exploit distributed execution across multiple nodes, where the dataset is sharded into independently processed partitions, thereby reducing memory requirements and computational cost per node. Nevertheless, the potential performance gain resulting from parallelisation is not necessarily linear with the number of nodes employed, as it depends on the resulting index topology. For example, for the NYTimes dataset, a centralised configuration using a radius of $r=1.03$ and an $np/gl$ ratio of 0.1 achieves a recall of 95.8\%, with an index-building time of 80.0~s and a search time of 0.93~s/query. Using the same configuration across 10 nodes results in a recall of 97.7\%, while reducing the index-building time to 34.7~s and the search time to 0.59~s/query.}

\section{CONCLUSIONS AND FUTURE WORK}
\label{sec:conclusion}

This paper presented PDASC, a distributed ANN search algorithm whose \modificacion{index design jointly addresses three concerns commonly treated separately in the ANN literature: support for arbitrary dissimilarity functions, native distributed execution, and compact per-node memory footprint. By limiting memory requirements and avoiding reliance on GPUs, TPUs, or other hardware-intensive optimisations, PDASC is well suited to resource-aware and energy-conscious deployments.}


Through an extensive structural sensitivity analysis, we showed that the resulting index topology, largely governed by the prototype-to-group ratio and the number of computing nodes, plays a central role in shaping the trade-off among recall, computational cost, and memory footprint. Empirical results across diverse datasets and distance functions demonstrate that PDASC consistently achieves competitive recall while requiring a smaller or comparable per-node index size than state-of-the-art ANN methods, \modificacion{making it a practical alternative for large-scale similarity search in memory-constrained environments.}



Several opportunities exist for further refinement and research. A natural direction is to improve the computational efficiency of the current implementation
. \modificacion{Another promising direction concerns adaptive and data-aware radius-selection strategies. In particular, dynamically refining the search radius during index traversal according to the local structure of the explored regions could enable PDASC to better exploit its inherent ability to adapt to heterogeneous densities and structural variations within the dataset.} Additionally, future work could focus on improving the selection of representative points to construct the hierarchical structure, for instance, by incorporating core-set techniques that capture the underlying data distribution more accurately.

\bibliographystyle{unsrtnat}
\bibliography{mybibfile}






\newpage
\section*{SUPPLEMENTARY MATERIAL}

\subsection*{1. Approximation quality and variability}
\label{sec:approximation-quality}

Table~\ref{tab:recall_algorithms} summarises the recall achieved by the evaluated algorithms across datasets and parameter configurations, complementing the NYTimes results presented in the main manuscript. Mean and SD denote the mean recall and its standard deviation, respectively, while Min and Max report the extreme values observed across the experiments. $P_{05}$ and $P_{95}$ correspond to the 5th and 95th percentiles of the recall distribution. The last column reports the percentage of experiments in which a recall of at least 90\% was achieved. The results reflect the behaviour of each method over the set of evaluated configurations, rather than the performance of a single selected configuration.

\begin{table}[h!]
\centering
\caption{Summary statistics of recall (\%) for the different algorithms across datasets.}
\label{tab:recall_algorithms}
\resizebox{0.8\columnwidth}{!}{
\begin{tabular}{llrrrrrrr}
\toprule
& & \multicolumn{6}{c}{Recall (\%)} & \\
\cmidrule(lr){3-8}
Dataset & Algorithm & Mean & SD & Min & $P_{05}$ & $P_{95}$ & Max
& Recall $\geq 90\%$ (\%) \\
\midrule
GLOVE
& ANNOY & 59.3 & 32.5 & 12.9 & 13.1 & 99.1 & 99.7 & 30.0 \\
& HNSW  & 48.1 & 32.8 & 0.0  & 0.2  & 90.9  & 95.2  & 5.9 \\
& IVF   & 88.8 & 20.5 & 33.9 & 39.8 & 100.0 & 100.0 & 77.4 \\
& LSH   & 53.2 & 36.0 & 0.0  & 0.0  & 89.9  & 91.1  & 5.1 \\
& PDASC & 97.2 & 4.37 & 77.2 & 87.8 & 100.0 & 100.0 & 92.4 \\
& PYNN  & 22.8 & 32.9 & 0.4  & 0.8  & 99.9  & 100.0 & 10.0 \\
\midrule
MNIST
& ANNOY & 95.9 & 5.46 & 83.7 & 85.3 & 100.0 & 100.0 & 78.6 \\
& HNSW  & 81.1 & 35.4 & 1.0  & 1.69 & 100.0 & 100.0 & 73.4 \\
& IVF   & 91.7 & 17.8 & 39.9 & 45.2 & 100.0 & 100.0 & 78.9 \\
& LSH   & 57.4 & 18.4 & 25.3 & 25.8 & 73.0  & 73.2  & 0.0 \\
& PDASC & 97.0 & 6.78 & 53.7 & 82.3 & 100.0 & 100.0 & 91.7 \\
& PYNN  & 48.9 & 31.7 & 13.0 & 16.6 & 100.0 & 100.0 & 18.6 \\
\midrule
MovieLens10M
& HNSW  & 74.4 & 33.7 & 0.7  & 1.2  & 99.0 & 99.1 & 53.7 \\
& PDASC & 94.7 & 5.75 & 75.8 & 83.6 & 99.2 & 99.6 & 81.2 \\
& PYNN  & 35.2 & 36.0 & 1.9  & 1.94 & 99.1 & 99.2 & 16.7 \\
\midrule
Municipalities
& PDASC & 99.6 & 1.30 & 89.8 & 98.1 & 100.0 & 100.0 & 99.1 \\
& PYNN  & 35.7 & 32.7 & 6.0  & 7.16 & 98.0 & 98.3 & 13.5 \\
\bottomrule
\end{tabular}
}
\end{table}

As can be observed, PDASC consistently achieves high mean recall across the
evaluated datasets, while maintaining comparatively low variability. ANNOY also provides high and stable recall on MNIST, whereas IVF generally achieves competitive mean recall but with greater variability. In contrast, HNSW, LSH and PYNN show substantially larger variations across the evaluated configurations, with lower average recall in several cases. These results suggest that, under the configurations considered in this study, PDASC provides a favourable balance between approximation quality and variability across parameter configurations.

\begin{figure}[h!]
    \centering
    \includegraphics[width=0.8\linewidth]{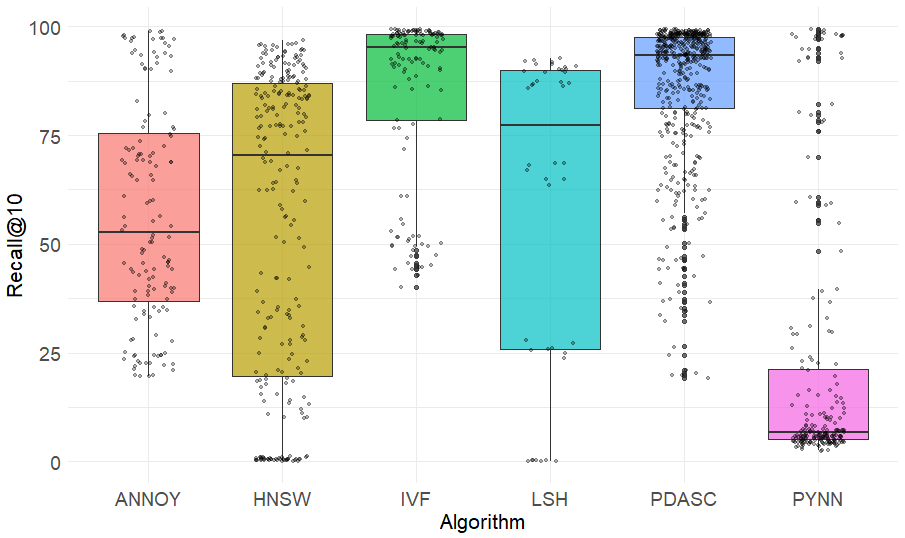}
    \caption{Distribution of recall on the NYTimes dataset for the evaluated
    ANN algorithms across different configurations.}
    \label{fig:global_recall}
\end{figure}

Fig.~\ref{fig:global_recall} presents the distribution of recall values obtained on the NYTimes dataset across different parameter configurations and experimental repetitions, providing a visual representation of the variability discussed in the main manuscript.

\subsection*{2. Influence of the $np/gl$ ratio on PDASC approximation quality}
\label{sec:ratio-recall}

Table~\ref{tab:recall_ratio} summarises the recall obtained for different
$np/gl$ ratios across all datasets. The results show a clear relationship between the $np/gl$ ratio and approximation quality across the evaluated datasets. In general, increasing the ratio leads to a decrease in mean recall and in the percentage of runs achieving at least $90\%$ recall. This decrease in recall is also accompanied by an increase in the standard deviation, indicating that larger ratios tend to produce not only lower recall but also less consistent results. Conversely, lower $np/gl$ ratios generally yield high recall with low variability, suggesting more accurate and stable performance. Although the magnitude of this effect varies across datasets, the overall trend is consistent: approximation quality and stability tend to deteriorate as the $np/gl$ ratio increases.

\begin{table}[h]
\centering
\caption{Summary statistics of recall (\%) for different PDASC $np/gl$ ratios across datasets.}
\label{tab:recall_ratio}
\resizebox{0.8\columnwidth}{!}{
\begin{tabular}{lrrrrrrrr}
\toprule
& & \multicolumn{6}{c}{Recall (\%)} & \\
\cmidrule(lr){3-8}
Dataset & Ratio & Mean & SD & Min & $P_{05}$ & $P_{95}$ & Max & Recall $\geq 90\%$ (\%) \\
\midrule
GLOVE
& 0.02 & 99.9 & 0.101 & 99.7 & 99.7 & 100.0 & 100.0 & 100.0 \\
& 0.05 & 99.7 & 0.574 & 98.2 & 98.3 & 100.0 & 100.0 & 100.0 \\
& 0.10 & 99.1 & 0.924 & 96.8 & 97.3 & 100.0 & 100.0 & 100.0 \\
& 0.20 & 98.0 & 2.020 & 93.8 & 93.9 & 99.9  & 100.0 & 100.0 \\
& 0.33 & 95.4 & 3.660 & 88.0 & 88.5 & 99.5  & 99.6  & 91.7 \\
& 0.50 & 90.8 & 5.970 & 77.2 & 81.6 & 97.9  & 98.5  & 62.5 \\
\midrule
MNIST
& 0.02 & 99.6 & 1.010 & 96.2 & 97.2 & 100.0 & 100.0 & 100.0 \\
& 0.05 & 99.1 & 1.560 & 94.8 & 96.1 & 100.0 & 100.0 & 100.0 \\
& 0.10 & 98.5 & 2.350 & 90.1 & 95.9 & 100.0 & 100.0 & 100.0 \\
& 0.20 & 97.5 & 3.520 & 89.5 & 90.6 & 100.0 & 100.0 & 95.8 \\
& 0.33 & 94.4 & 8.020 & 70.8 & 78.4 & 100.0 & 100.0 & 79.2 \\
& 0.50 & 89.7 & 12.800 & 53.7 & 65.1 & 99.9 & 100.0 & 66.7 \\
\midrule
MovieLens10M
& 0.02 & 98.2 & 1.860 & 94.1 & 94.6 & 99.5 & 99.6 & 100.0 \\
& 0.05 & 97.3 & 2.350 & 93.5 & 93.7 & 99.2 & 99.2 & 100.0 \\
& 0.10 & 96.1 & 3.310 & 89.8 & 89.8 & 99.1 & 99.1 & 87.5 \\
& 0.20 & 94.5 & 4.690 & 85.8 & 87.5 & 99.0 & 99.1 & 81.2 \\
& 0.33 & 93.2 & 5.680 & 84.3 & 84.9 & 98.8 & 98.8 & 68.8 \\
& 0.50 & 89.0 & 8.700 & 75.8 & 75.9 & 98.1 & 98.3 & 50.0 \\
\midrule
Municipalities
& 0.02 & 100.0 & 0.000 & 100.0 & 100.0 & 100.0 & 100.0 & 100.0 \\
& 0.05 & 100.0 & 0.000 & 100.0 & 100.0 & 100.0 & 100.0 & 100.0 \\
& 0.10 & 100.0 & 0.101 & 99.6  & 99.8  & 100.0 & 100.0 & 100.0 \\
& 0.20 & 99.7  & 0.534 & 98.0  & 98.8  & 100.0 & 100.0 & 100.0 \\
& 0.33 & 99.5  & 0.911 & 96.8  & 97.8  & 100.0 & 100.0 & 100.0 \\
& 0.50 & 98.2  & 2.940 & 89.8  & 92.4  & 100.0 & 100.0 & 93.8 \\
\midrule
NYTimes
& 0.02 & 96.9 & 2.210  & 90.3 & 92.7 & 99.2 & 99.4 & 100.0 \\
& 0.05 & 94.3 & 5.420  & 77.3 & 83.0 & 99.0 & 99.3 & 83.3 \\
& 0.10 & 93.6 & 6.380  & 75.7 & 77.9 & 98.9 & 99.3 & 83.3 \\
& 0.20 & 87.9 & 10.800 & 57.0 & 62.0 & 98.3 & 99.0 & 56.0 \\
& 0.33 & 78.7 & 16.000 & 37.2 & 46.3 & 97.1 & 97.7 & 33.3 \\
& 0.50 & 66.4 & 22.400 & 19.2 & 21.5 & 96.5 & 97.1 & 14.3 \\
\bottomrule
\end{tabular}
}
\end{table}

\end{document}